# VERY HIGH RESOLUTION LAND COVER MAPPING OF URBAN AREAS AT GLOBAL SCALE WITH CONVOLUTIONAL NEURAL NETWORKS

T. Tilak [1*], A. Braun [1], D. Chandler [1], N. David [1], S. Galopin [1], A. Lombard [2], M. Michaud [1], C. Parisel [1], M. Porte [1], M. Robert [1]

[1] IGN, Institut National de l'Information Géographique et Forestière, 73, avenue de Paris 94165 Saint-Mandé France - (thomas.tilak, arnaud.braun, david.chandler, nicolas.david, sylvain.galopin, michael.michaud, camille.parisel, matthieu.porte, marjorie.robert)@ign.fr

[2] CEREMA, Centre d'études et d'expertise sur les risques, l'environnement, la mobilité et l'aménagement, 1 avenue du Colonel Roche 31400 Toulouse France - amelie.lombard@cerema.fr

**ABSTRACT:**

This paper describes a methodology to produce a 7-classes land cover map of urban areas from very high resolution images and limited noisy labeled data. The objective is to make a segmentation map of a large area (a french department) with the following classes: asphalt, bare soil, building, grassland, mineral material (permeable artificialized areas), forest and water from 20cm aerial images and Digital Height Model.
We created a training dataset on a few areas of interest aggregating databases, semi-automatic classification, and manual annotation to get a complete ground truth in each class.
A comparative study of different encoder-decoder architectures (U-Net, U-Net with Resnet encoders, Deeplab v3+) is presented with different loss functions.
The final product is a highly valuable land cover map computed from model predictions stitched together, binarized, and refined before vectorization.

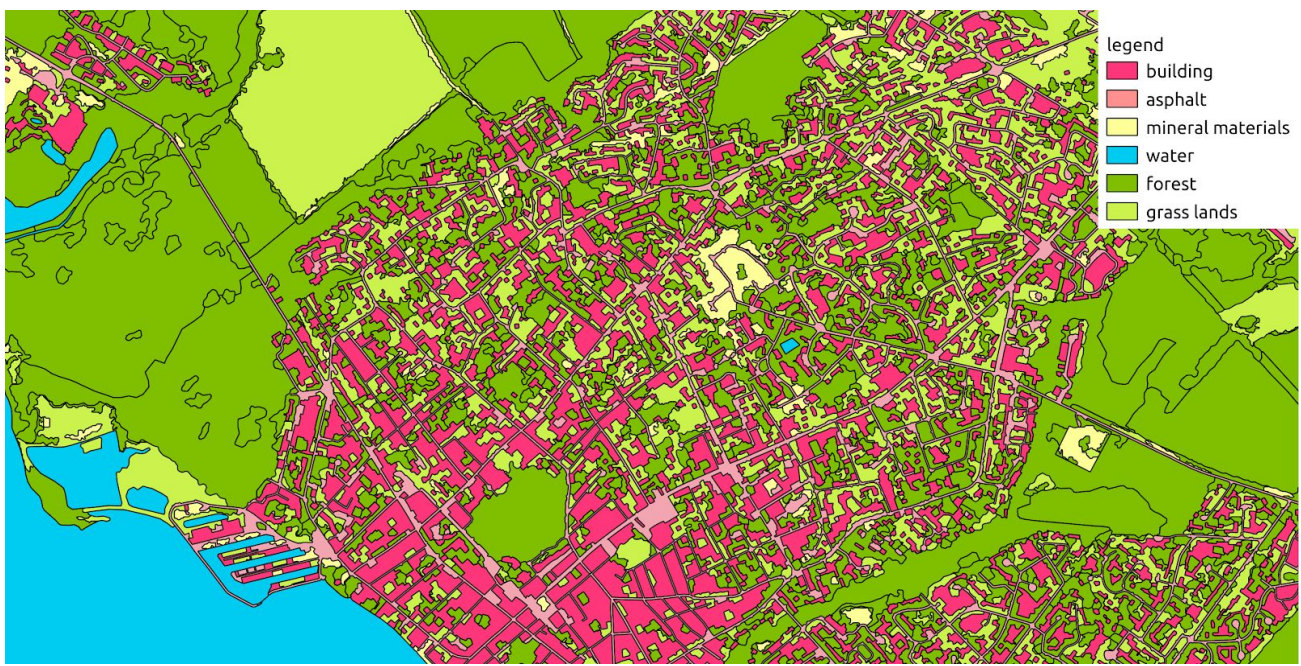

Figure 1. Final land cover map, Andernos-les-Bains, France

* Thomas Tilak thomas.tilak@ign.fr, 6, avenue de l'Europe 31520 Ramonville

# 1. INTRODUCTION

Land cover maps are cartographic products widely demanded by land managers. As they provide a quantification of land resources into thematic categories (e.g. forest, water or asphalt surface), land use/land cover maps are used to measure current conditions and how they are changing.

The different classes composing the final land cover map are derived from OCSGE product (Description of OCSGE). They have been selected to best fit the description of urban areas: asphalt, bare soil, building, grassland, mineral material, forest, and water. The final product (Figure 1) can be used for deriving environmental indicators and measuring land consumption.

In the following paper, we propose a full methodology to produce a very high resolution 7-classes land cover map at a large scale with aerial images. Specific attention has been given to subsequent tasks:
- creating a consistent and representative training dataset for each class
- benchmarking different models and configurations
- producing a relevant land cover map from model predictions

As databases (BDTOPO® or OpenStreetMap®) do not describe all the objects of each class, manual annotation is needed. We chose to create full ground truth on specific areas of interest (AOI) in the production zone to facilitate the labelling process. In the end, only 10% of the targeted zone is used for training.
Semantic segmentation of aerial images with convolutional neural networks is then performed from this training dataset.

Since the publication of an encoder-decoder deep convolutional network architecture (Ronneberger et al., 2015), scientists have tried to improve semantic segmentation on medical images (Ibtehaz, Rahman, 2019) or terrestrial images (Kaiming He et al., 2015; Liang-Chieh Chen et al., 2018) and others have transposed those solutions to earth observation with high resolution images taken from satellite (Nagesh Kumar Uba, 2019), airplane (Ce Zhang et al., 2019) or drone (Zhang et al, 2019). Multiple competitions have been recently carried out to improve semantic segmentation results on earth observation images with a provided dataset (Spacenet challenges, ISPRS semantic labelling contest). Lots of work on roads or building extraction from satellite images (Zhengxin Zhang et al., 2017) have been published.

We propose to reuse top-class semantic segmentation architectures (U-Net with ResNet encoders, DeeplabV3+) on very high resolution aerial images.
This article also describes some post-processing techniques to convert model predictions to a relevant land cover map.

# 2. METHODOLOGY

## 2.1 Input data

We chose a sample size of 512 pixels for convenience as we are using a DeeplabV3+ model with a receptive field of 470 pixels and a GPU with limited memory (8Gb).

**2.1.1 Images**: As input data, we use a mosaic of aerial images with a resolution of 20cm. The camera has 4 bands (Red, Green, Blue and Near-Infrared) and original pixel depth is reduced to 8 bits before the mosaicing process. As a complement, a Digital Surface Model - derived from the aerial images by photogrammetric techniques - is used in combination with a Digital Terrain Model to produce a Digital Height Model (also called normalized DSM) that we concatenate as a 5th channel to input images.

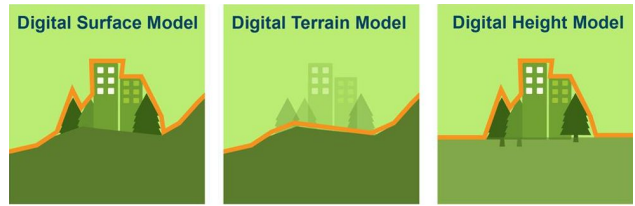

Figure 2. Digital Height Model (source Intermap on Twitter)

**2.1.2 Masks:** Associated masks are created separately for each class.
Some classes are labelled with different techniques detailed in paragraph 3.1.
A multi-channel binary image is produced to store the label information with a band for each class.

**2.1.3 Samples:** In the end we feed the model with a tuple of (image, mask) with the shapes:

```
image.shape = (512, 512, 5)
# 5 for (R, G, B, IR, MNH)
mask.shape = (512, 512, 8)
# 8 for the 7-classes + complementary mask
```

Figure 3 shows a representation of a training sample.

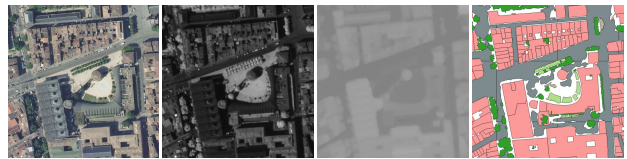

Figure 3. Example of input data (RGB, Infra-Red, DHM, labels)

## 2.2 Network architectures

Three encoder-decoder network configurations are assessed in this article. Original U-Net is the baseline configuration and we measure precisely improvements brought by more sophisticated architectures: U-Net with ResNet50 as encoder on one side and DeeplabV3+. Those models have very different characteristics and specificities.
- U-Net is a simple encoder-decoder convolutional network. The encoder part processes the input image through several convolutional and down-sampling layers, giving semantically rich feature maps. The decoder part consists of convolutional and upsampling layers combined with skip connections with the encoder feature maps to access further spatial information. Those connections furthermore alleviate the risk of vanishing or exploding gradients.

- ResNet, used as an encoder, differs mainly by the use of skip connections between consecutive layers in the encoder path, allowing to learn residual feature maps.
- DeepLabV3+ (Figure 5) puts together several developments in Deep Convolutional Neural Networks (Liang-Chieh Chen et al, 2018). The network follows an encoder-decoder scheme, using Atrous Convolution and Atrous Spatial Pyramid Pooling to extract rich multi-scale information and Depthwise convolution to reduce computational costs. The model is implemented with MobileNetv2 (Sandler et al, 2018) as backbone.

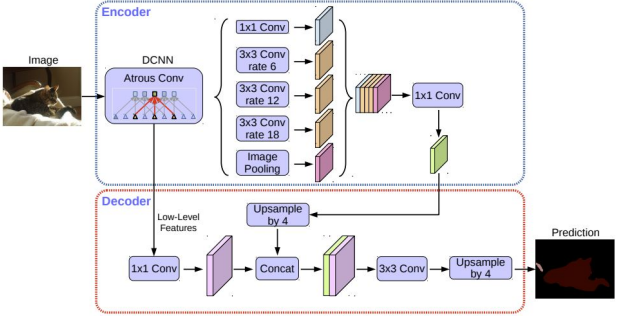

Figure 4. Deeplab V3+ architecture

Those models differ not only in their architecture but also in the number of parameters, impacting both the amount of training data and of computation time required. The smallest models are based on U-Net and the largest on Resnet18 encoders.

### 2.3 Loss functions

Several loss functions are evaluated in this work.
- Cross-Entropy (CE) loss takes into account the variation between prediction and target. But it is evaluated on individual pixels, thus large objects contribute more to it than small ones which can be an issue when dealing with land cover objects.
- Weight-balanced Cross-Entropy (WCE) loss is used to tackle the problem of an unbalanced training dataset. Under-represented classes get higher weights to assure that all classes contribute equally to the loss value.
- Binary Cross-Entropy (BCE) is used to evaluate the benefits of allowing the network to perform a multi-label classification task (each pixel could belong to multiple classes).
- A combo loss with BCE + Jaccard. Jaccard loss is computed from the Jaccard index defined by:

$$J(y_{true}, y_{pred}) = \frac{\sum(y_{true} * y_{pred})}{\sum(y_{true} y_{pred}) - \sum(y_{true} * y_{pred})} \quad (1)$$

with $y_{pred}$ a predicted segmentation and $y_{true}$ a ground truth labelling.
Jaccard index is transformed to obtain Jaccard loss:

$$J_l(y_{true}, y_{pred}) = 1 - J(y_{true}, y_{pred}) \quad (2)$$

The combo loss is a weighted combination of BCE and Jaccard loss:

$$Combo_{loss}() = 0.25 * BCE() + 0.75 * J_l() \quad (3)$$

Using Jaccard loss mixed with BCE is a quite common technique used to handle imbalanced input.

### 2.4 Post-processing

The raw result is a probability map or heat map with a band per class. Each pixel is represented by a vector of belonging probability with values rescaled between 0 and 255.
We perform a 2-level thresholding:
- binarization of heat map with a low threshold
- vectorization
- averaging of probability by object
- filtering of objects whose average probability is below a high threshold

A last, a refining process is made to remove objects according to their size, then geometries are simplified.

## 3. EXPERIMENTS

We focus on producing a land cover map of the urban area of Gironde department (Figure 7). Aerial images taken by the French Mapping Agency in 2018 with a resolution of 20cm, the associated Digital Surface Model and BDALTI® are used to build the image samples.

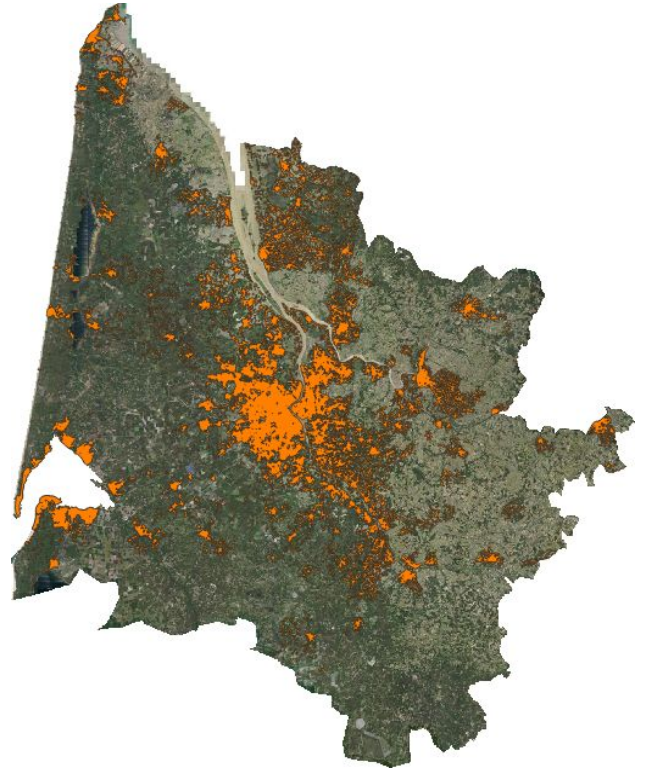

Figure 5. Urban areas of Gironde department

### 3.1 Training dataset elaboration

The dataset specifications are issued from some practical choices during its production.
Most of the time, when training datasets are not provided for experiments, they are generated by picking a large number of random samples among the targeted area. We decided instead to create our dataset focusing on 5 consistent areas of interest carefully selected to best represent the overall territory

characteristics (Figure 6). This choice has been determined by two main reasons:
- It is easier to automatize or semi-automatize the training label production, with standard machine learning tools, on large images tiles than on many small patches.
- It brings more flexibility in the choice of relevant size and resolution for training samples.

**3.1.1 AOIs characteristics:** This dataset is mainly dedicated to urban area segmentation.

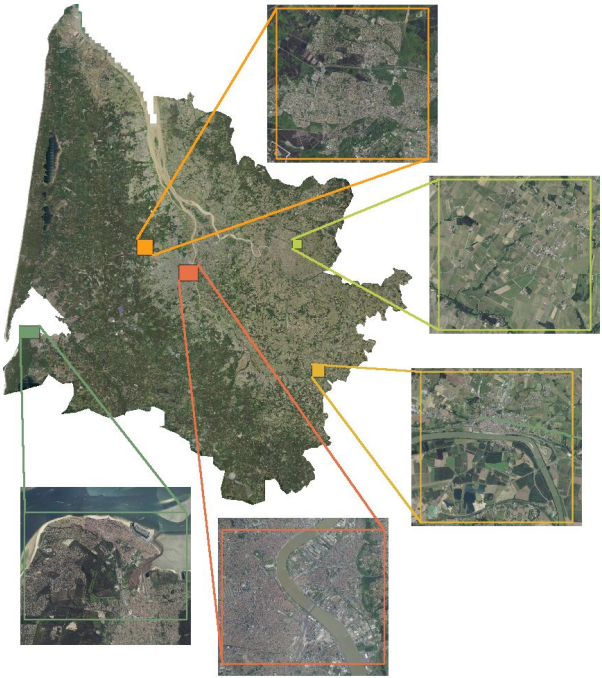

Figure 6. Training zones

So, areas where land cover is mainly forest, natural or cropland have not been taken into consideration. But in order to catch urban area diversity, territories with different urban morphology have been manually selected. The chosen urban morphology is coastal and touristic areas (dark green in Figure 6), dense downtown (red in Figure 6), residential suburban zones (orange in Figure 6) and small towns in agricultural territories (light green and yellow in Figure 6).

**3.1.2 Labelling:** Each class is labelled separately. French BDTOPO® database provides consistent information for buildings, roads, natural water areas. But it does not contain any parking lots or sidewalks useful for asphalt class nor urban grassland or forests and mineral materials. Those objects are extracted either with manual annotation or semi-automatic techniques from classical machine learning. Some of them are detailed in the following sections.

**Classical machine learning for vegetation:** To build the training dataset for vegetation classes (forest and grassland), classical machine learning is used. We combine a pixel-based classification through random forests and a large-scale mean shift segmentation.
Vegetation indices NDVI and NDWI, as well as RGB and Near-Infrared bands, are used as input of a supervised random forest classifier. We finally combine the pixel classification with regions of a large scale mean shift segmentation. Thanks to a majority voting on pixel class among each region, we get larger homogenous regions. Taking advantage of DHM and NDVI to detect vegetation and trees, the results are acceptable as is for ground truth.
All those processings are done using functionalities of orfeo-toolbox (Grizonnet et al., 2017) command-line tools.

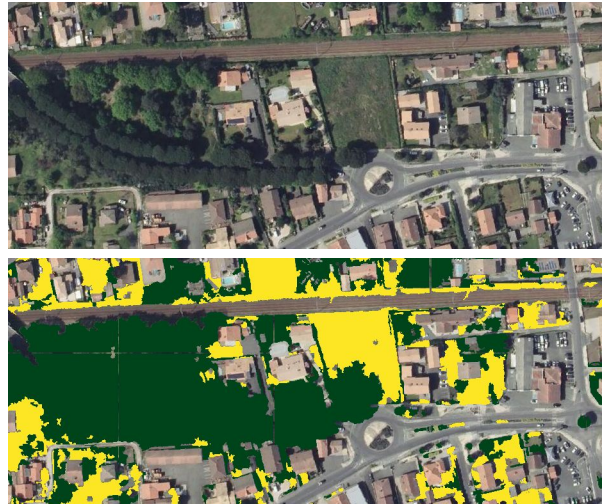

Figure 7. Example of forest (green) and grassland (yellow) ground truth generated semi-automatically

**CNN predictions for mineral materials areas:** Minerals objects (Figure 8) are not easy to extract, as they have no signature in DHM or radiometric bands.

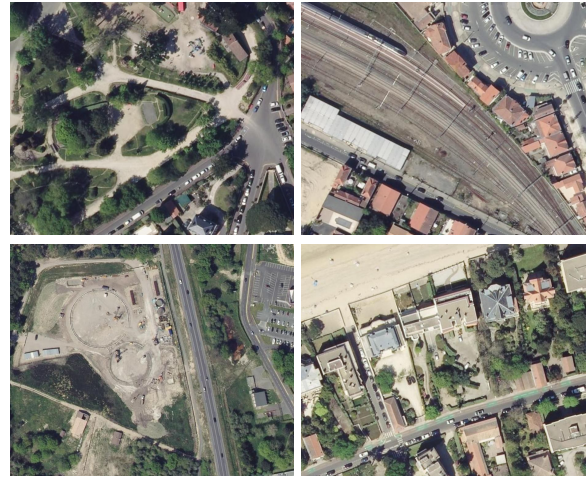

Figure 8. Example of mineral materials objects (path, railways, dumps, yard)

The strict definition for mineral material class is: every area with a permeable but man-made coverage, including path, railways, yards, … It can be easily mistaken for asphalt or concrete. The common approach to this problem relies on massive manual annotation for each training zone. We rather choose to use a basic U-Net trained on a tiny subset of the training data manually labelled (5 km$^2$ over 100 km$^2$). The network predictions over the 5 training areas are then post-processed (vectorization, simplification, merging with available vector data like road network) and manually corrected to constitute a ground truth.

**3.1.3 Training dataset consistency:** From each AOI, we extract square samples of 512 pixels width.

The constituted training dataset is evaluated to ensure that each class is adequately represented.

Here are few statistics on the dataset showing that it is quite balanced with only 2 classes under-represented: mineral materials and bare soil (see Table 2).

| stat | value |
|---|---|
| share_multilabel | 0.02 |
| entropy | 0.98 |
| nb_class | 4.72 |
| nb_samples | 7777 |

Table 1. Global statistics of the dataset

Table 1 gives global statistics with:
- *share_multilabel*: Overall share of pixels with multiple classes.
- *nb_class*: Mean of the number of classes present in each sample
- *entropy*: Mean of the class distribution entropy for each sample. For each sample, the entropy is at least 0 if a single class is represented and at most log(C) with C the number of classes. The entropy of a sample is log(C) if every class is equally represented. The greater the entropy, the semantically richer the sample is.
- *nb_samples*: the number of samples in the dataset.

| class | pixel_frequency |
|---|---|
| building | 0.12 |
| forest | 0.28 |
| grassland | 0.2 |
| asphalt | 0.12 |
| mineral materials | 0.02 |
| bare soil | 0.01 |
| water | 0.1 |
| unlabelled | 0.17 |

Table 2. Per class statistics

### 3.2 Benchmarking configurations

The dataset is split in train|test|validation parts:
- training set represents 64% of the total
- validation set (16% of the total) used between each epoch to prevent overfitting and decide when convergence is reached
- test set (20%) is used to provide the metrics reported in the following paragraphs

The experiment is run on a desktop computer with the following characteristics: Intel(R) Core(TM) i7-6950X CPU @ 3.00GHz and 16Gbits of RAM and a GPU NVIDIA GeForce RTX 2070 with 8Gbits of RAM.

The source code of the processing pipeline is based on pytorch and will be made available soon.

Training hyper-parameters are:
- training on 300 epochs with a patience of 20 (if loss do not decrease on the validation set for 20 epochs, the training is stopped)
- learning rate starts at 0.001 and is halved when the loss on validation does not decrease
- Adam optimizer

Every network is trained from scratch.

We use basic data augmentation with a randomized 90-degree rotation.

**3.2.1 Configuration performances:** The measure used to compare configurations is the Intersection Over Union (IOU) index also known as the Jaccard index described by formula 1. This index measures the overlap between prediction and ground truth. It is preferred to pixel accuracy as it is not affected by class imbalance. The following table presents IOU results (in percentage) computed on the test dataset which has 1935 samples.

| model | loss | building | forest | grass | asphalt | mineral | bare soil | water |
|---|---|---|---|---|---|---|---|---|
| U-Net | CE | 77.76 | 84.14 | 74.56 | 77.68 | 40.81 | 72.82 | 96.12 |
| U-Net | BCE | 77.9 | 84.74 | 75.23 | 77.42 | 41.20 | 74.91 | 95.85 |
| U-Net | WCE | 72.99 | 80.39 | 69.08 | 62.48 | 20.50 | 57.86 | 93.04 |
| U-Net | Combo | 77.60 | 84.75 | 74.80 | 77.56 | 39.44 | 72.12 | 95.69 |
| Resnet18 | CE | 77.27 | 84.75 | 73.67 | 71.62 | 21.67 | 36.96 | 93.24 |
| Resnet18 | BCE | 76.60 | 84.06 | 72.31 | 71.07 | 21.64 | 9.990 | 93.09 |
| Resnet18 | WCE | 76.55 | 83.40 | 72.37 | 69.85 | 27.26 | 32.73 | 94.31 |
| Resnet18 | Combo | 77.27 | 85.27 | 73.61 | 71.12 | 21.38 | 36.67 | 94.38 |
| Deeplab V3+ | CE | **82.66** | 85.83 | 77.78 | **79.36** | **46.59** | 82.57 | 96.43 |
| Deeplab V3+ | BCE | 82.32 | 86.95 | 78.42 | 77.98 | 41.85 | 86.77 | 97.46 |
| Deeplab V3+ | WCE | 81.78 | 83.97 | 77.26 | 76.63 | 41.57 | **89.54** | 96.75 |
| Deeplab V3+ | Combo | 82.49 | **87.47** | **78.87** | 77.70 | 45.23 | 77.35 | **97.59** |

Table 3. IOU in percentage for each class and architecture configuration

The overall result is quite good even if we note a severe defect for mineral material class which is known to be problematic. The training dataset is not very reliable for this class and it is a real challenge even for a human eye to distinguish between concrete and mineral material on proposed images.

**3.2.2 Analysis:** Although it has more trainable parameters, Resnet18 based models perform equally with simple U-Net or even worse for difficult classes (mineral materials and bare soil). It seems that our dataset is too small to be able to benefit from the complexity of Resnet18.

| | number of parameters | training time/epoch |
|---|---|---|
| U-Net | 453 393 | 9 minutes |
| Deeplab V3+ | 6 236 784 | 11 minutes |
| Resnet18 | 24 507 344 | 16 minutes |

Table 4. Global training figures

DeeplabV3+ architecture with its atrous convolution feature performs better than U-Net based models.

Loss performances are quite equivalent for this experiment. We note that the use of Binary Cross Entropy degrades measured performances and combo loss (BCE + Jaccard loss) improves IOU for many classes but to a relatively limited extent.

**3.2.3 Visual comparison:** Results are generally equivalent, though we can point out some minor differences. Deeplab models delineate objects more precisely than U-Net and give better performance on small objects.

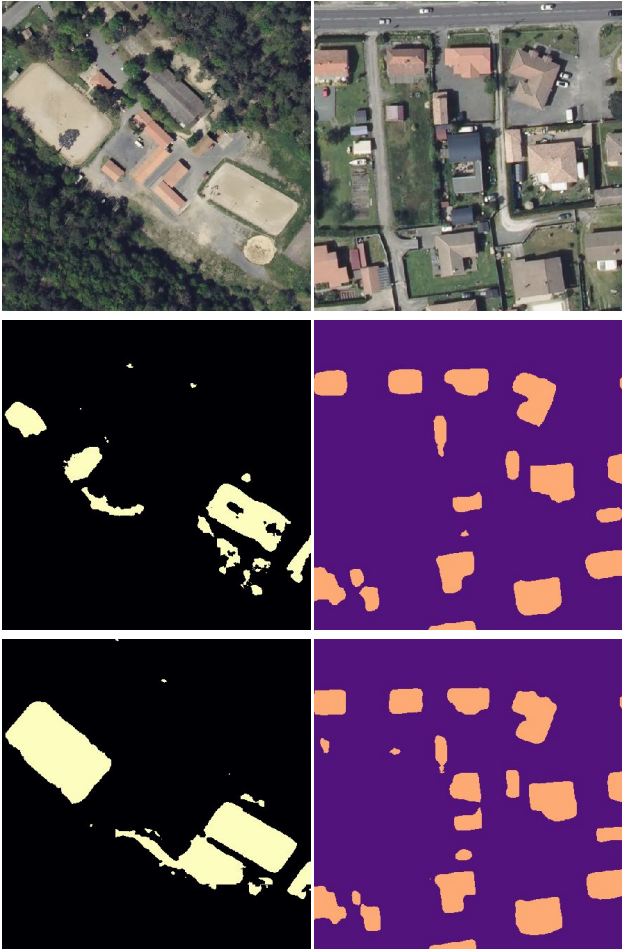

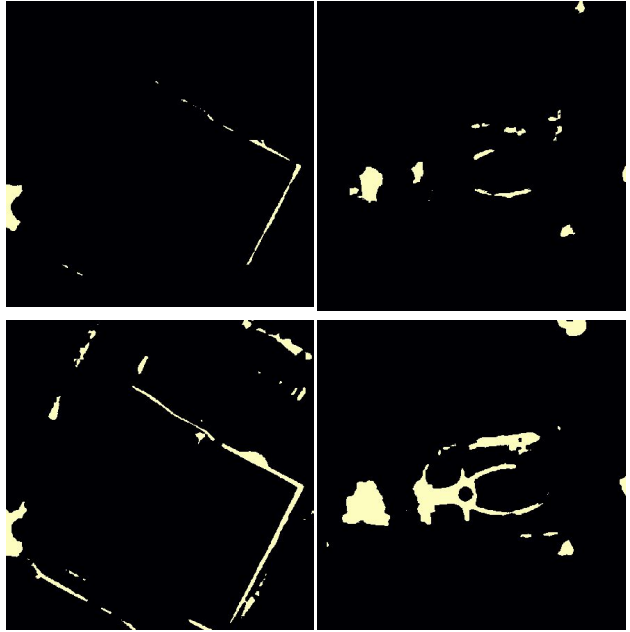

Figure 9. Comparison of U-Net and Deeplab binarized predictions (top: RVB images, middle: U-Net predictions and bottom: Deeplab predictions) for mineral materials and forest classes.

Figure 9 shows in the first column dirt pitches being segmented more precisely with Deeplab and in the second column building being extracted by Deeplab model while ignored by U-Net.

Whereas visual differences between CE, BCE or even Combo losses are not significant, WCE loss performs better for subsidiary classes. Mineral materials or bare soils segmentation get improved.

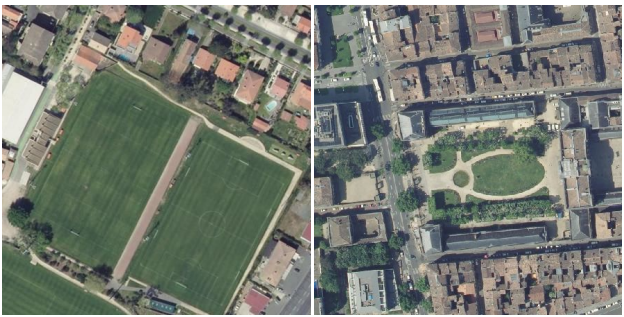

Figure 10. Comparison of BCE and WCE losses binarized predictions (top: RVB images, middle: BCE predictions and bottom: WCE predictions) for mineral materials and building classes.

### 3.3 Global land cover map from model predictions

Having a trained model performing well on test images is one thing many studies would settle for. But the final purpose here is to produce a global vectorized land cover map describing elementary objects.

**3.3.1 From model predictions to stitched map:** A trained model predicts probability of class belonging for tiled images also called patches. Stitching those predicted tiles together is challenging. Vignetting effects can appear on predictions due to a poor consideration of models' receptive fields. As (Bohao Huang et al, 2018) points out, modern Convolutional Neural Networks used in this article do not guarantee translational equivariance due to zero-padding and strides operations underlying.

We present results of 2 methods to perform a smooth combination of predictions:
- clipping edges of patches
- blending predictions with a windowing function

**Clipping edges of patches:** The following predictions are made on larger tiles (2048 x2048 pixels) to optimize the available GPU memory . A fishnet is calculated with an overlap size chosen with respect to the network receptive field.

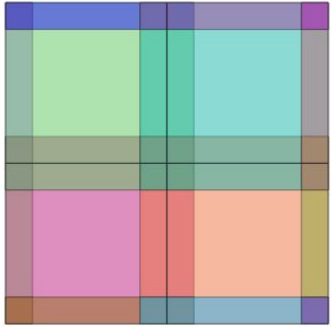

Figure 11. Overlapping tiles for prediction

The resulting map is an image where the connections between the tiles are barely visible.

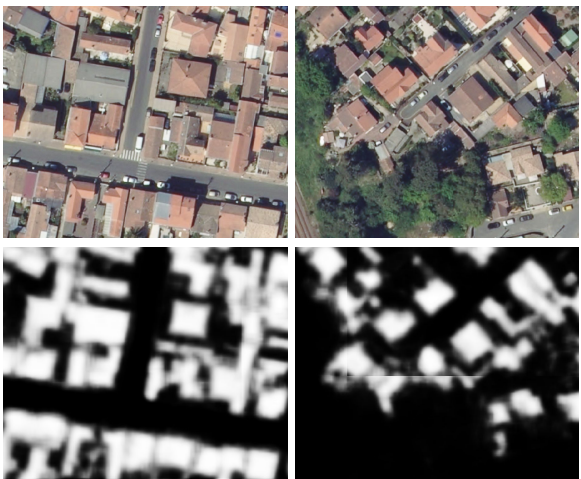

Figure 12. Extract of stitched raw predictions for building class

**Blending prediction with a windowing function:** Work from (Chevalier, 2017) has been reused to improve the final result. The algorithm makes several predictions of a single tile with transformations applied (rotations, mirrors) and averages them. Then it uses a 2d spline function to blend overlapping predictions together.

The result is of very high quality. Not only each pixel prediction is much more precise, but connections between prediction tiles have totally disappeared.

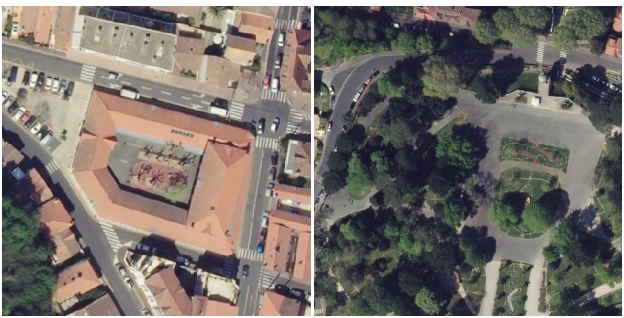

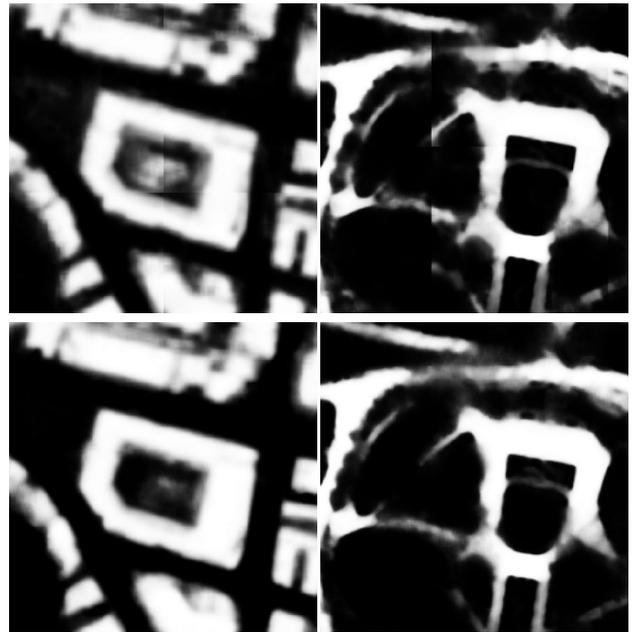

Figure 13. Extract of predictions for building class (left column) and asphalt (right column) with edge clipping (middle row) and smooth blending (bottom row)

**Vectorized land cover map:** Figure 1 shows an extract of the final product build upon predictions map from DeeplabV3+ model with BCE loss.

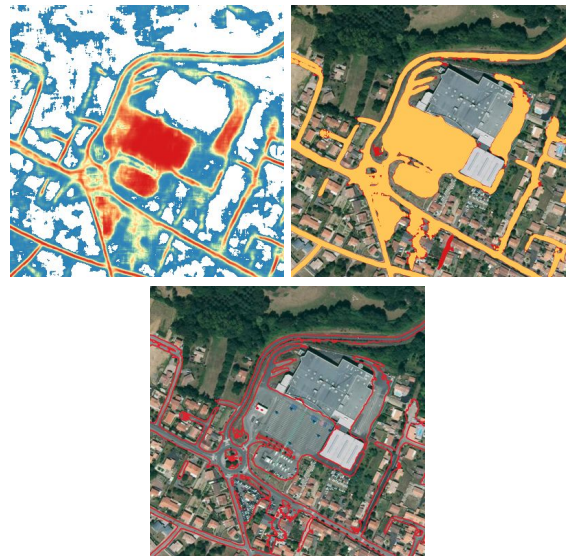

Figure 14. Vectorization of asphalt class, from left to right and top to bottom: heat map from predictions, comparison of 2 double-threshold processes (yellow and red), vector

The second image in Figure 14 shows how we can adjust the thresholds to retrieve more objects from predictions. The final result is refined with a removal of elements of small size and geometry simplification.

## 4. CONCLUSIONS

This article gives hints to build a training dataset for classes asphalt, bare soil, buildings, grassland, mineral materials

(permeable artificialized areas), forest and water from scratch when full manual annotation is not an option.

DeeplavV3 architecture shows very good results for semantic segmentation of aerial images.

Dataset quality is the most important criteria to obtain good results on prediction. Prediction concatenation is a real challenge when a seamless heat map is expected.

This experiment has been conducted within a project whose objective is modernizing production lines for land cover products on French territory. Land cover maps produced on Gironde department are then derived to a generalized product and will be used by public authorities to monitor the land consumption.

In the future, the dataset will be completed with new images captured at other dates and in other departments (with different lighting conditions). We will study how the model can be generalised with more data augmentation techniques dealing with radiometry.

The current experiment focuses on urban areas but there is a work in progress to generate land cover maps on rural areas with slightly different classes though.

## 5. ACKNOWLEDGEMENTS

This project is built upon the research work carried out by Tristan Postadjian (Postadjian et al, 2017). His experience in using neural networks to extract information from images has been very helpful to build the whole pipeline. We could also thank researchers from IGN laboratory LASTIG, especially Sebastien Giordano and Arnaud Le Bris.

## 6. REFERENCES

Description of OCSGE, IGN, French Mapping Agency. OCSGE OCcupation du Sol à Grande Échelle https://artificialisation.biodiversitetousvivants.fr/bases-donnees/ocs-ge

Ronneberger, O., Fischer, P., Brox, T., 2015: U-Net: Convolutional Networks for Biomedical Image Segmentation U-Net: Convolutional Networks for Biomedical Image Segmentation, https://arxiv.org/abs/1505.04597v1

Ibtehaz, N., Rahman, M. S., 2019: MultiResUNet : Rethinking the U-Net Architecture for Multimodal Biomedical Image Segmentation, https://arxiv.org/abs/1902.04049

Kaiming He, Xiangyu Zhang, Shaoqing Ren, Jian Sun, 2015: Deep Residual Learning for Image Recognition, https://arxiv.org/abs/1512.03385

B. Zhang, Y. Kong, H. Leung and S. Xing, 2019: Urban UAV Images Semantic Segmentation Based on Fully Convolutional Networks with Digital Surface Models, Tenth International Conference on Intelligent Control and Information Processing (ICICIP)

Liang-Chieh Chen, Yukun Zhu, George Papandreou, Florian Schroff, Hartwig Adam, 2018: Encoder-Decoder with Atrous Separable Convolution for Semantic Image Segmentation, https://arxiv.org/abs/1802.02611

Ce Zhang, Sargent, I., Xin Panc, Huapeng Lid, Gardiner, A., Haree, J., Atkinson, P.M., 2019: Joint Deep Learning for land cover and land use classification, https://doi.org/10.1016/j.rse.2018.11.014

SpaceNet challenges, SpaceNet LLC, https://spacenet.ai/

ISPRS semantic labelling contest, ISPRS, 2018, http://www2.isprs.org/commissions/comm3/wg4/semantic-labeling.html

Zhengxin Zhang, Qingjie Liu, Yunhong Wang, 2017: Road Extraction by Deep Residual U-Net, https://arxiv.org/abs/1711.10684

Nagesh Kumar Uba, 2019: Land Use and Land Cover Classification Using Deep Learning Techniques, https://arxiv.org/abs/1905.00510

Tsung-Yi Lin, Priya Goyal, Girshick, R.,, Kaiming He, Dollár, P., 2017: Focal loss for dense object detection, https://arxiv.org/abs/1708.02002

Grizonnet, M., Michel, J., Poughon, V., Inglada, J., Savinaud, M., Cresson, R., 2017: Orfeo ToolBox: Open source processing of remote sensing images, Open Geospatial Data, Software and Standards.

Sandler, M., Howard, A., Menglong Zhu, Zhmoginov, A., Liang-Chieh Chen, 2018: MobileNetV2: Inverted Residuals and Linear Bottlenecks, https://arxiv.org/pdf/1801.04381

Bohao Huang, Reichman D., Collins, L.M., Bradbury, K., Malof, J. M., 2018: Tiling and Stitching Segmentation Output for Remote Sensing: Basic Challenges and Recommendations https://arxiv.org/abs/1805.12219

Postadjian, T., Le Bris, A., Sahbi, H., and Mallet, C., 2017: Investigating the Potential of Deep Neural Networks for Large-Scale Classification of Very High Resolution Satellite Images, ISPRS Ann. Photogramm. Remote Sens. Spatial Inf. Sci., IV-1/W1, 183–190, https://doi.org/10.5194/isprs-annals-IV-1-W1-183-2017, 2017

Chevalier, G., 2017: Smoothly-Blend-Image-Patches, https://github.com/Vooban/Smoothly-Blend-Image-Patches